\documentclass[11pt]{article}
\usepackage{graphicx}
\usepackage{multirow}
\usepackage{float} 
\usepackage{url,tabularx,amsfonts,slashed,amsmath,array}
\usepackage{lscape}

\begin{document}
\begin{flushright}
preprint SHEP-12-24\\
\today
\end{flushright}
\vspace*{1.0truecm}
\begin{center}
{\large\bf Distinguishing $Z'$ models with polarised top pairs}\\
\vspace*{1.0truecm}
{\large L. Basso$^1$, K. Mimasu$^2$ and S. Moretti$^{2,3}$}\\
\vspace*{0.5truecm}
{\it $^1$Physikalisches Institut, Albert-Ludwigs-Universit\"at Freiburg\\
D-79104 Freiburg, Germany}\\
\vspace*{0.25truecm}
{\it $^2$School of Physics \& Astronomy, University of Southampton, \\
Highfield, Southampton, SO17 1BJ, UK}\\
\vspace*{0.25truecm}
{\it $^3$Particle Physics Department\\ Rutherford Appleton Laboratory\\
Chilton, Didcot, Oxon OX11 0QX, UK}\\
\end{center}

\vspace*{0.5truecm}
\begin{center}
\begin{abstract}
\noindent
We study the sensitivity of top pair production at the Large Hadron Collider (LHC) to the nature of an underlying $Z'$ boson, including full tree level standard model background effects and interferences. We demonstrate that exploiting combinations of asymmetry observables will enable one to distinguish between a selection of `benchmark' $Z'$ models while assuming realistic final state reconstruction efficiencies and error estimates.
\end{abstract}
\end{center}

\section{Introduction}\label{sec:intro}
$Z'$ bosons are a ubiquitous feature of theories beyond the Standard Model (SM) appearing from a variety of concepts such as $U(1)$ gauge extensions of the SM motivated by supersymmetry or grand unified theories, Kaluza-Klein excitations of SM gauge fields or excitations of composite exotic vector mesons in technicolor theories to name a few.

The obvious channel to search for such objects at hadron colliders is Drell-Yan (DY) production of a lepton pair, i.e., $pp(\bar p) \to (\gamma,Z,Z') \to \ell^+\ell^-$, where $\ell=e,\mu$. The strongest limits on $Z'$'s at both Tevatron and the LHC come from this signature. The Tevatron places limits on the $Z'$ mass, $M_{Z'}$, at around 1 TeV~\cite{Tevatron} (for a sequential $Z'$) while the latest LHC limits lie around 2.3 TeV~\cite{LHC}. Phenomenological studies on how to measure the $Z'$ properties and couplings to SM particles in this clean DY channel have been performed.

These proceedings summarise a recently published paper~\cite{Basso:2012sz} addressing the use of the top-antitop final state, i.e., $pp(\bar p)\to (\gamma,Z,Z') \to t\bar t$, to probe these $Z'$ properties. While it may not have as much `discovery' scope as the DY channel, owing to the large QCD background combined with the complex six-body final state and the associated poor reconstruction efficiency, it remains important to extract the couplings of new physics to the top quark. Furthermore, the fact that the top decays before hadronising, transmitting spin information to its decay products, allows for the definition of spin asymmetry observables which provide an extra handle on $Z'$ couplings not present in non-decaying final states. 

We study the scope of the LHC to profile a $Z'$ boson mediating $t\bar t$ production, in both standard kinematic variables as well as spatial/spin asymmetries, by adopting some benchmark scenarios for several realisations of the sequential, Left-Right symmetric and $E_6$ based $Z'$ models (specifically, the same as those in Accomando \emph{et al.}~\cite{Accomando:2010fz}). Specifically, the issue of distinguishability of various models using these observables is addressed.
\section{Asymmetries and $Z'$ couplings}\label{sec:defcalc}
We define the asymmetry observables considered with the aim of determining their power to discriminate between $Z'$s. We refer the reader to our paper for a more detailed discussion on these as well as the selection of benchmark models, statistical uncertainties and definitions of significance. This study investigated charge (spatial) and spin asymmetries and their dependence on top couplings to profile and distinguish the models considered. 

A selection of charge asymmetry variables were investigated with the most sensitive found to be $A_{RFB}$, defined by the rapidity difference of the top and antitop, $\Delta y=|y_t|-|y_{\bar{t}}|$, while also cutting on the boost of the $t\bar{t}$ system. This increases the contribution from the $q\bar{q}$ initial state by probing regions of higher partonic momentum fraction, $x$, where its parton luminosity is more important:
\small
\begin{align}
    A_{RFB}&=\frac{N(\Delta y > 0)-N(\Delta y < 0)}{N(\Delta y > 0)+N(\Delta y < 0)}\Bigg |_{|y_{t\bar{t}}|>|y^{cut}_{t\bar{t}}}.
\end{align}
\normalsize
The two spin asymmetries considered, termed double ($LL$) and single ($L$), are defined as follows:
\small
\begin{align}\label{asy_ALL}
A_{LL}=\frac{N(+,+) + N(-,-) - N(+,-) - N(-,+)}{N_{Total}}\quad;\quad A_{L}=\frac{N(-,-) + N(-,+) - N(+,+) - N(+,-)}{N_{Total}}
\end{align}
\normalsize
where $N$ denotes the number of observed events and its first(second) argument corresponds to the helicity of the top (anti)quark.

Defining a generic neutral current interaction in terms of gauge, vector and axial couplings $g^{\prime}$, $g_{V}$ and $g_{A}$ with the Feynman rule
\begin{equation}
i\frac{g^{\prime}}{2}\gamma^{\mu}(g_{V}-g_{A}\gamma^{5}),
\end{equation}
\noindent
the asymmetric term in the polar angle of the $Z'$ matrix element (which contributes to the charge asymmetry) is proportional to $g^i_V g^i_A g^t_V g^t_A$ where $i$ labels the initial state partons. The dependence on the chiral couplings of the spin asymmetries can be expressed analytically, using helicity formulas from Arai \emph{et al.}~\cite{Arai:2008qa} (also derived independently with the guidance of Hagiwara \emph{et al.}~\cite{Hagiwara:1985yu}): 
\begin{eqnarray}\label{analytic_ALL}
A_{LL}^{i} &\propto& \Big( 3\, (g^t_A)^2\beta^2 + (g^t_V)^2(2+\beta^2)\Big) \, \Big( (g^i_V)^2 + (g^i_A)^2 \Big) \, ,\\ \label{analytic_AL}
A_{L}^{i}  &\propto& g^t_A\, g^t_V \,\beta\,\Big( (g^i_V)^2 + (g^i_A)^2 \Big) \, ,
\end{eqnarray}
for a neutral gauge boson exchanged in the $s$-channel, where $\beta=\sqrt{1-4\, m_t^2/\hat{s}}$. These imply that $A_{LL}$ depends only on the square of the couplings similarly to the total cross section and that $A_{L}$ is only non-vanishing for non-zero vector and axial couplings of the final state tops and is additionally sensitive to their relative sign.
\section{Results}\label{sec:results}
We present a selection of results profiling the spatial and spin asymmetry distributions of the benchmark $Z^{\prime}$ models compared to the SM including interference effects. The set of benchmarks are split into two categories: those with a vanishing vector or axial coupling (the $E_{6}$ models with the `B-L' generalised left-right symmetric model) are classed as the `$E_{6}$' type while the rest, with both couplings non-zero, are referred to as the `generalised models'. The variables described in section~\ref{sec:defcalc} were computed as a function of the $t\bar{t}$ invariant mass within $\Delta M_{t\bar t}=|M_{Z^{\prime}}-M_{t\overline{t}}|<500$ GeV and compared to the tree-level SM predictions. The code exploited for our study is based on helicity amplitudes, defined through the HELAS subroutines~\cite{HELAS}, and built up by means of MadGraph~\cite{MadGraph}. CTEQ6L1~\cite{cteq} Parton Distribution Functions (PDFs) were used, with the factorisation/renormalisation scale at $Q=\mu=2m_t$. VEGAS~\cite{VEGAS} was used for numerical integration.
\begin{figure}[h!]	
\centering
\includegraphics[width=0.32\linewidth]{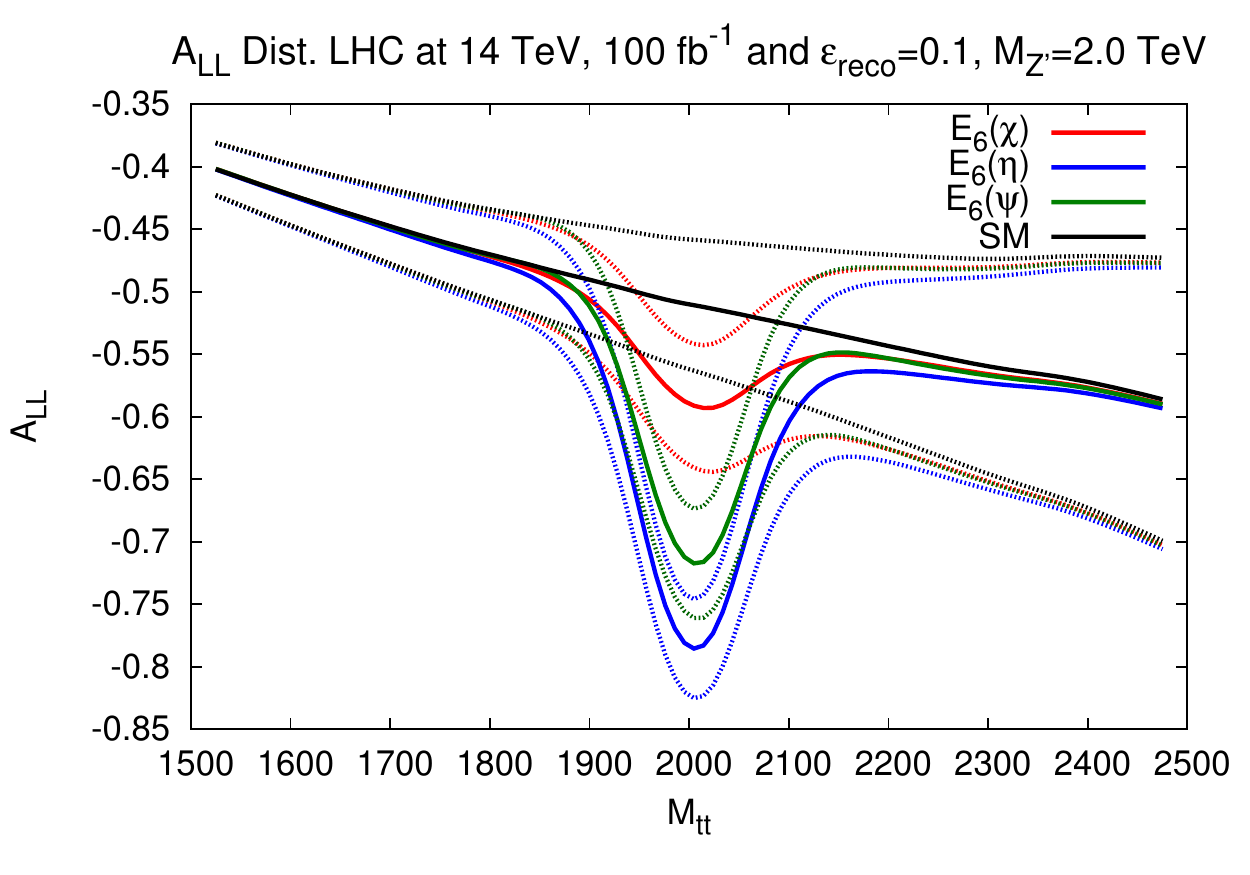}
\includegraphics[width=0.32\linewidth]{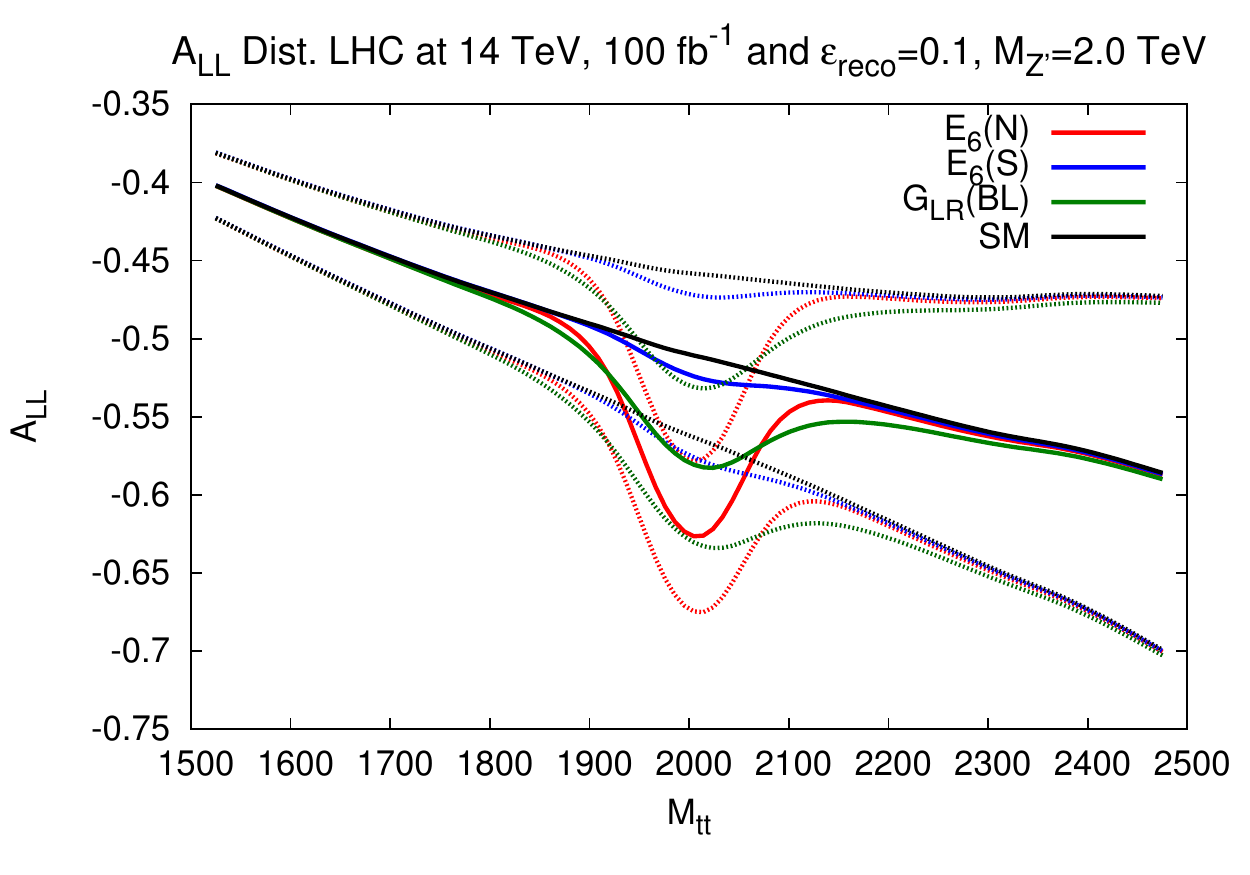}
\includegraphics[width=0.32\linewidth]{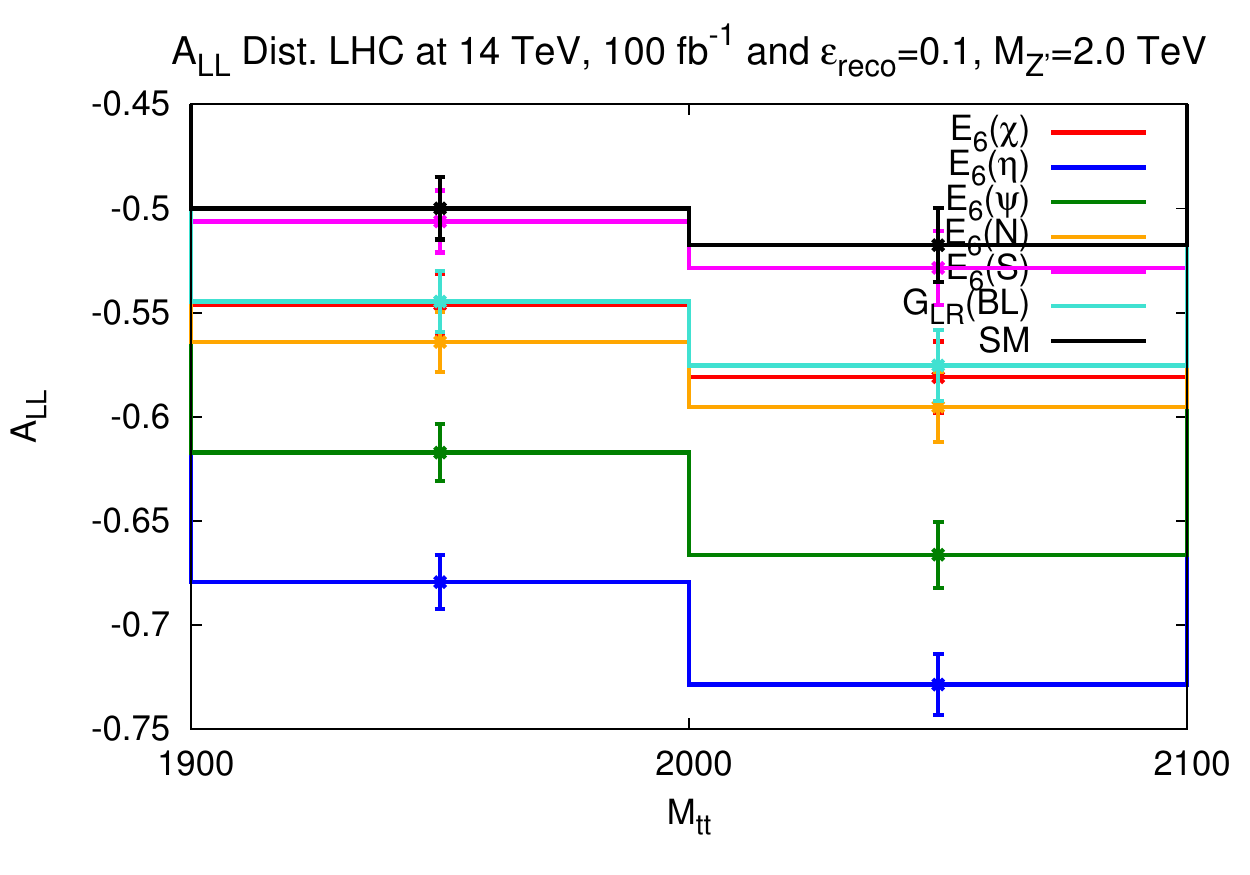}\\
\includegraphics[width=0.32\linewidth]{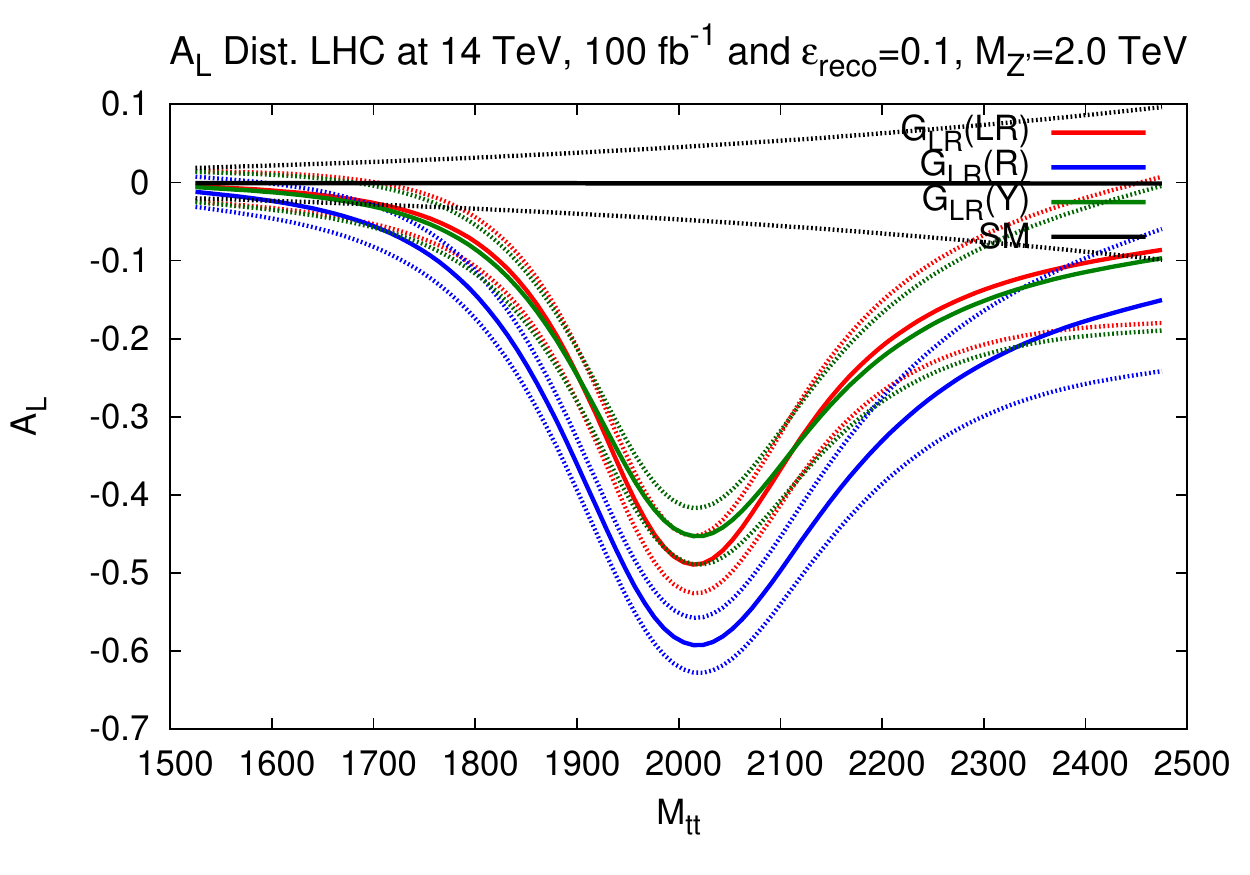}	\includegraphics[width=0.32\linewidth]{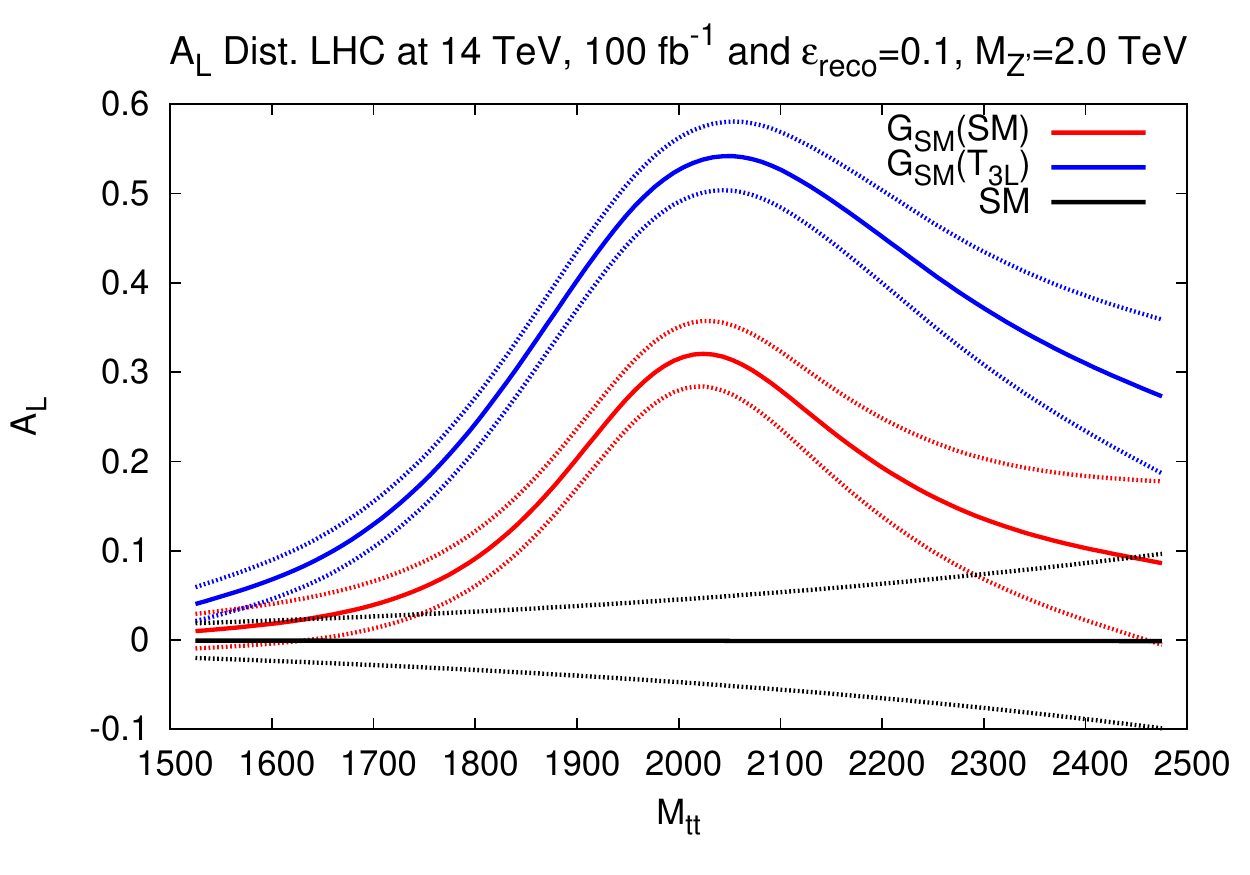}	\includegraphics[width=0.32\linewidth]{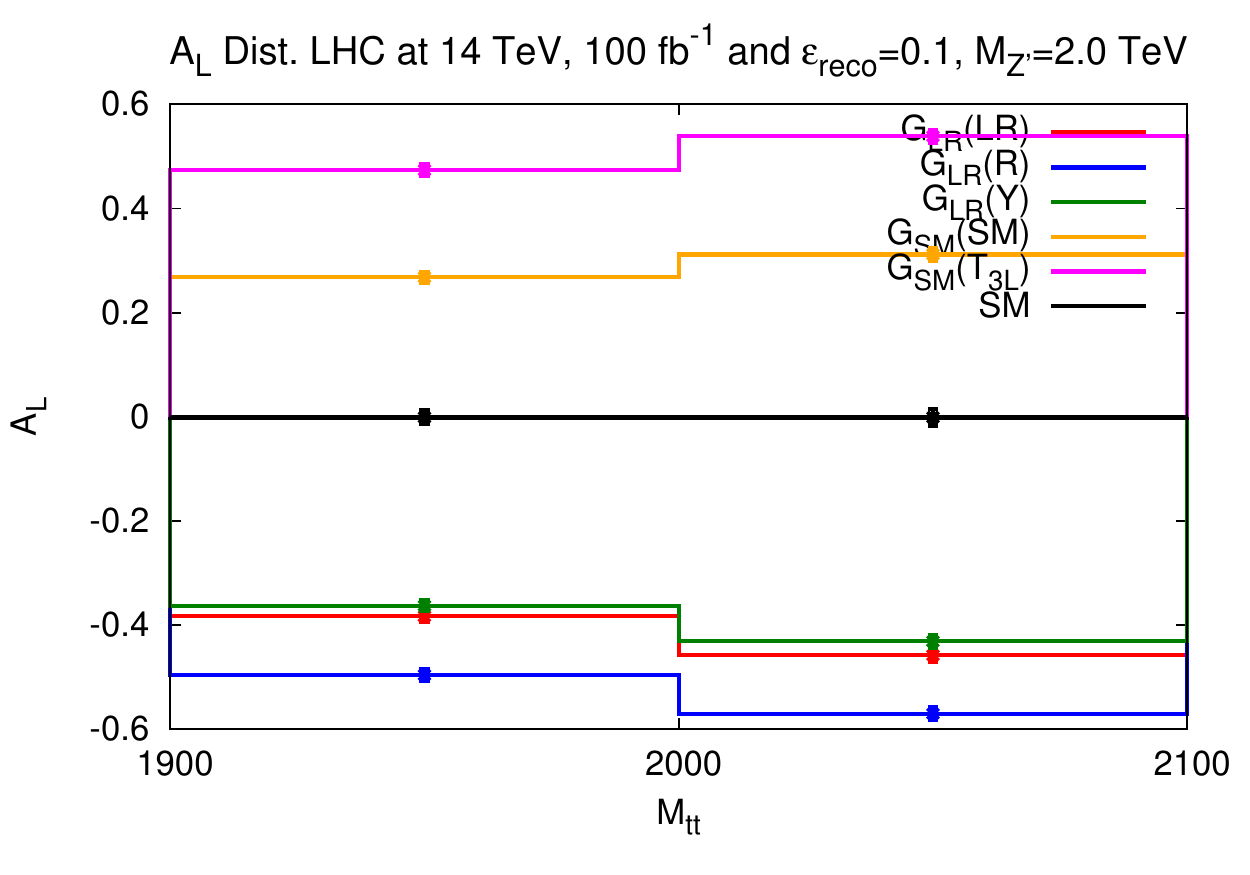}
\caption{$A_{LL}$ for the $E_{6}$-type models (\emph{upper}) and $A_{L}$ for the generalised models (\emph{lower}) binned in $M_{t\overline{t}}$ with $M_{Z^{\prime}}$=2 TeV for the LHC at 14 TeV assuming 100 fb$^{-1}$ of integrated luminosity. Rightmost plots show the distribution in two 100 GeV bins either side of the $Z^{\prime}$ peak. Statistical uncertainties are shown assuming a bin width of 50 GeV compatible with typical experimental resolutions.}\label{fig:asys}
\end{figure}

The example $A_{LL}$ and $A_{L}$ distributions in figure~\ref{fig:asys} show visible effects around the $Z'$ peak including statistical uncertainties and folding in an estimated 10\% reconstruction efficiency of the $t\bar{t}$ pair assuming the use of all decay channels. Systematic uncertainties may also be important but would require a study beyond parton level. We also show a selection of two bin plots integrating the cross sections over an `on-peak' range ($\Delta M_{t\bar t}<100$ GeV) and evaluating the corresponding partially integrated asymmetry. Overall the figures (particularly the two bin plots) show that the majority of benchmark models can be distinguished from one another using these variables, noting in particular the sensitivity of $A_{L}$ to the relative sign of the vector and axial couplings which allows for a clear distinction between the $G_{SM}$ (sequential) and $G_{LR}$ models (left-right symmetric) models in the `generalised' category. $A_{LL}$ depends on the couplings in the same way as the total cross section and therefore models that cannot be distinguished in the invariant mass spectrum will remain so in this variable. Table~\ref{tab:signif_LHC14_ARFB} is an example of the statistical studies made where the significance of an observable (in this case $A_{RFB}$) is examined assuming 100 fb$^{-1}$ of integrated luminosity. Here, the significance between various models is a measure of how well they can be distinguished. In most cases the significances are important, showing that this variable would certainly be effective in disentangling the `generalised' models with the discrimination decreasing slightly for higher masses. However we do demonstrate in the paper that the $A_{L}$ variable is more effective to this end although the observable itself may be more difficult to extract. Although not shown in these proceedings, a similar statistical analysis was performed in determining how much integrated luminosity would be required to achieve a significance of 3 between models in the various observables. We showed that in many cases, the models can be distinguished at the relatively early stages of the LHC ($\sim$ 100 fb $^{-1}$ at 14 TeV). The cases where this is not possible reflect mostly instances where the couplings are too similar and would be difficult to disentangle.
\begin{table}[h!]
\centering
\small
\begin{tabular}{|c|c|c|c|c|c|c|}\hline
$A_{RFB}$      &$SM$&$G{LR}(LR)$ &$G{LR}(R)$&$G{LR}(Y)$&$G{SM}(SM)$&$G{SM}(T_{3L})$\\ \hline
$SM$         & --       & 9.2(3.3) & 12.8(5.7) & 8.6(3.4) & 5.2(2.2) & 12.2(7.2)  \\
$G{LR}(LR)$ & 4.8(2.2) & --        & 4.1(2.5)  & 0.8(0.1)  & 4.9(1.2) & 3.4(3.9) \\
$G{LR}(R)$  & 6.4(3.6) & 1.9(1.5)  & --        & 4.9(2.4)  & 9.2(3.7)& 0.8(1.4) \\
$G{LR}(Y)$  & 4.4(2.2)&0.6($\ll$ 1)& 2.5(1.5)  & --        & 4.1(1.3) & 4.1(3.9) \\
$G{SM}(SM)$ & 2.6(1.4) & 2.7(0.8)  & 4.7(2.4)  & 2.2(0.9)  & --       & 8.4(5.2) \\
$G{SM}(T_{3L})$& 5.9(4.2) & 1.4(2.1)  & 0.4(0.7)  & 2.0(2.2)  & 4.2(3.0) &  --    \\
\hline
\end{tabular}
\normalsize
\caption{Significance for $A_{RFB}$ values around the $Z^{\prime}$ peak of generalised models, for the LHC at 14 TeV only. Upper triangle for $M_{Z'}=2.0$ TeV and lower triangle for $M_{Z^{\prime}}$=2.5 TeV. Figures refer to $\Delta M_{t\bar t}<100$(500) GeV.}\label{tab:signif_LHC14_ARFB}
\end{table}
\section{Conclusion}\label{sec:summary}
We have presented an overview of a phenomenological study on classes of $Z^{\prime}$ models in both spin and spatial asymmetries of $t\bar{t}$ production and showed that there is much scope to observe deviations from the SM and even distinguish between various models, particularly for spin asymmetries. This suggests that the $t\bar{t}$ channel would certainly be a useful complement to the more popular DY channel in the aim of profiling a $Z'$ resonance should one be observed in the near future.

It is worth noting that the classes of models studied are benchmarks put forward to set bounds on $Z^{\prime}$ masses best probed in the di-lepton channels. Other models could be better suited to the $t\bar{t}$ channel, such as leptophobic/top-phillic $Z^{\prime}$s occurring in composite/multi-site and extra-dimensional models. The profiling techniques discussed in this study would be increasingly more applicable in these scenarios.
\section*{Acknowledgments}
The work of KM and SM is partially supported through the NExT Institute. LB is supported by the Deutsche Forschungsgemeinschaft through the Research Training Group grant GRK\,1102 \textit{Physics of Hadron Accelerators}. We would like to thank E. Alvarez for pointing out the discussion on systematic uncertainties.

\end{document}